\documentclass[runningheads]{llncs}

\usepackage{amssymb}
\usepackage{graphicx}

\title{A Simple Framework to Typify Social Bibliographic Communities}
\author{Christoph Schommer\institute{University of Luxembourg, Dept. of Computer Science and Communication (CSC), MINE Research Group @ ILIAS Laboratory, 6 Rue Coudenhove Kalergi, 1359 Luxembourg, Luxembourg. Email: christoph.schommer @ uni.lu}}

\begin{document}

\maketitle
\begin{abstract}
Social Communities in bibliographic databases exist since many years, researchers share common research interests, and work and publish together. A social community may vary in type and size, being fully connected between participating members or even more expressed by a consortium of small and individual members who play individual roles in it. In this work, we focus on social communities inside the bibliographic database \textit{DBLP} and characterize communities through a simple typifying description model. Generally, we understand a publication as a transaction between the associated authors. The idea therefore is to concern with directed associative relationships among them, to decompose each pattern to its fundamental structure, and to describe the communities by expressive attributes. Finally, we argue that the decomposition supports the management of discovered structures towards the use of adaptive-incremental mind-maps.
\end{abstract}

\section{Bibliographic Libraries}
\textit{DBLP} (\cite{ley:dblp}, \cite{ley:reu}) is an online database system regarding bibliographic entries of scientific publications inside Computer Science. It maintains publications of single authors or a set of authors, it is public and offers a retrieval interface to query publication entries. Since March 2008, more than one million entries have been added to the database. The first publications are of 1936, the latest of 2008. An intelligent DBLP core gathers data from known conferences, collecting entries from electronic publications. Interpreting each publication as a data stream element (the authors share some time together, converse, and produce a result), then this refers to similar problems and challenges given above. 

We have performed association discovery of DBLP entries throughout on a yearly basis for the last 72 years. This is because the data set has a certain size and is therefore more expressive than a data set on a monthly basis, but more precise than a less fine-granulated discretisation. The calculation of association rules has been done by using a sample in a way that we firstly set the \textit{item set frequency} threshold to 0.1\% and secondly the \textit{bayesian probability value} threshold to 5\%. Thirdly, since the number of received rules has still been tremendously high, we filtered some of them by taking the \textit{lift} - representing the proportion of the bayesian probability and the statistically independent case of an association rule $A\Rightarrow B$. In this respect, the number of scientific contributions had been downsized; we got a distribution curve having the pike in 1972. Although the publication of a scientific paper has a delay of a several months (submission date, evaluation period, publication), we counted each scientific paper to the year it has appeared. Following the yearly calculation of associative rules - beginning from 1936 and ending in 2007 - a various number of patterns occurred.

\begin{figure}[htbp]
   \centering
    \includegraphics[width=12.5cm]{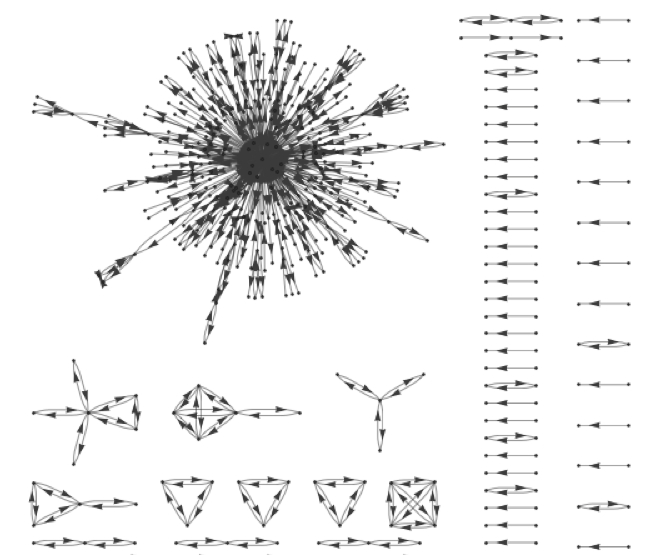}
    \caption{DBLP: Snapshot visualization of publishing communities (1994).}
   \label{fig:ximg-14}
\end{figure}

Some of them are presented in Figure \ref{fig:ximg-14}, showing the distribution of the publishing authors in the year of 1994. The publishing communities in 1994 are mainly characterized by a central \textit{star} consisting of many author nodes. Generally, the observations we have done, produced a diverse number patterns that are often quite similar and that base on very simple geometric structures. Most interestingly, patterns went away but appeared again, they stayed stable or disappeared forever. For example, the big star (Figure \ref{fig:ximg-14}) has not been existing before 1955, but appeared several times afterwards, for example in 1994, disappeared temporarily, and appeared again in 2006 (\textit{visiting pattern}). Simple structures like in Figure \ref{fig:ximg-1} are present continuously (\textit{constant pattern}), without any temporal break: this sounds interesting in the first moment, but in fact, it simply reflects a certain kind of noise within the set of results.

\section{A Simple Typifying Model}

Following our observations, we typify each associative pattern to their fundamental structure, and - since these structures are evocative of chemical basic modules - we label them in almost the same manner.

Each author node $i$ corresponds to an \textit{atomic author nucleus}, owning a certain activation $act_i$ and a number of atomic bonds with other nuclei. In the following model description, we keep these bonds unvalued although the strengthen between the adjacent atomic author nuclei exists per se.

\begin{figure}[htbp]
   \centering
   a) \includegraphics[width=4.5cm]{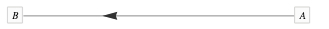}
   b) \includegraphics[width=4.5cm]{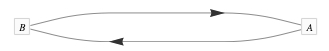} 
   \caption{Single atomic bonds (left) and double atomic bonds (right) between two atomic author nucleus.}
   \label{fig:ximg-1}
\end{figure}

As presented in Figure \ref{fig:ximg-1}, we call such a connection a \textit{single atomic bond}, as it describes the relationship between two \textit{author nuclei} on the lowest level, being single-directed. Furthermore, we call a relationship a \textit{double atomic bond}, if the relationship between two authors is undirected (see Figure \ref{fig:ximg-1}) in a sense that both the association between author nuclei A and B and B and A, respectively, are frequent. Both structures define the \textit{fundamental molecular structure} between two adjacent \textit{nuclei}. In case, that two \textit{nuclei} share two \textit{single atomic bonds} of different direction, we call this undirected relationship a \textit{bridge}. All these molecules are called \textit{2-ary} since only two \textit{atoms} are involved.  

\begin{figure}[htbp]
   \centering
   a) \includegraphics[width=6cm]{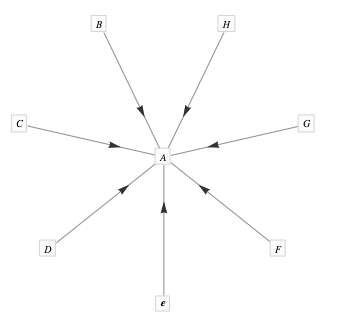}  \\
   b) \includegraphics[width=6cm]{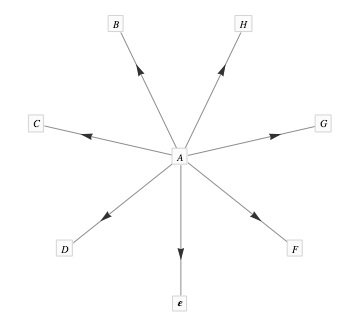} \\
   c) \includegraphics[width=6cm]{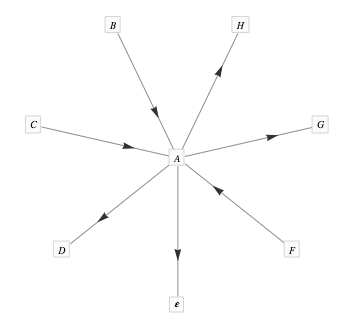} 
      \caption{2-ary types of molecule stars of different roles; the last one is a combination of both.}
   \label{fig:ximg-4}
\end{figure}

As for this, any combination of atomic structures results in further \textit{molecules}, being of different granularity and size. In general, we understand molecules as expressive in respect to their arity. Some examples of 2-ary \textit{fundamental molecules} are presented in Figure \ref{fig:ximg-4}. On the left, an atomic nucleus $A$ is being arranged as a centre of $k$ adjacent nuclei. We call this molecule structure a \textit{molecule star}, meaning that the inner nuclei is dependent on each adjacent nuclei: following the meaning of the bond, a publication of $A$ has always been done under the condition that someone else has published. In case that a \textit{author nucleus} is connected to only disjunctive \textit{nuclei} sharing no other \textit{bonds}, then the \textit{author nucleus} is still being 2-ary, otherwise n-ary.

For Figure \ref{fig:ximg-4}, we observe on the right side a \textit{molecule star} is the opposite of the previous one as $A$ is the originator of any publication, being the condition that others publish. With this, we differentiate between different roles of a \textit{nucleus}:

\begin{itemize}
    \item An \textit{author nucleus} A is a \textit{Trigger} if A influences another author B: $A\Rightarrow B$
    \item An \textit{author nucleus} A is a \textit{Reactor} if A is is influenced by another author B: $B\Rightarrow A$
    \item Two \textit{author nuclei} A and B define a \textit{bridge} if they share a \textit{double atomic bond}.
\end{itemize}

\begin{figure}[htbp]
   \centering
   a) \includegraphics[width=5cm]{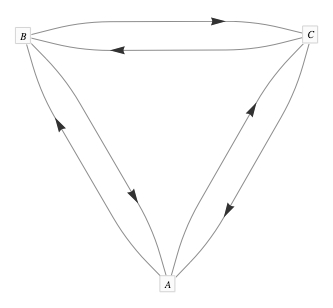}
   b) \includegraphics[width=5cm]{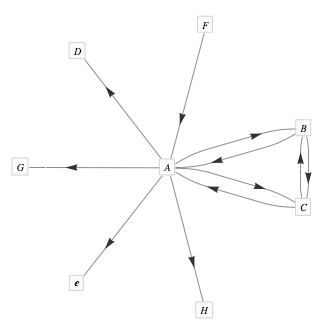}\\ 
   c) \includegraphics[width=5cm]{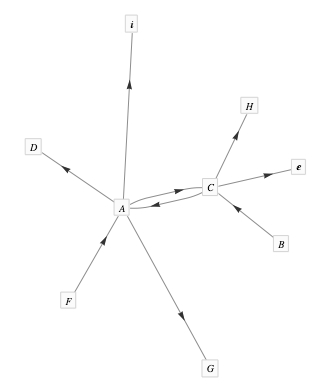} 
   d) \includegraphics[width=5cm]{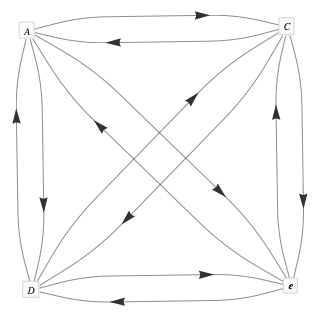}
      \caption{Selected molecular forms with molecular diamond (above, left) and molecular mixture of star and diamond (above, rights), mixture with single and double atomic bonds and molecular bridge (below, left), double diamond (below, right).}
   \label{fig:ximg-2}
\end{figure}

Some molecules structure representing a mixture of \textit{single} and \textit{double atomic bonds} are shown in Figure \ref{fig:ximg-2}. The structures a) and d) are fully composed of \textit{double atomic bonds}, whereas b) and c) share \textit{single atomic bonds} as well. We call the molecular structure in a) a \textit{molecular diamond} and the central sub-structure in c) a \textit{molecular bridge}. The structure in b) is a mixture of a \textit{molecular star} and a \textit{molecular diamond}, the structure in d) four overlapping \textit{molecular diamonds}. With this, \textit{molecular stars} can be seen communities that consist of an arbitrary number of triggers and reactors; and a \textit{molecular diamond} is nothing else than a composition of bridges. Furthermore, all \textit{molecules} are still 2-ary.

In this respect, a community is a set of \textit{author nuclei} that are connected by \textit{molecular structures}. For example, if $A\Rightarrow B$, then the \textit{nuclei} A and B are connected, where B is reachable by A. A collection of associations like $A\Rightarrow B$, $B\Rightarrow C$, and $C\Rightarrow A$ therefore yields on a structure where each author nuclei can be reached by another one.  Two disjunctive molecules define disjunctive social communities.

\begin{figure}[htbp]
   \centering
    a)\includegraphics[width=5cm]{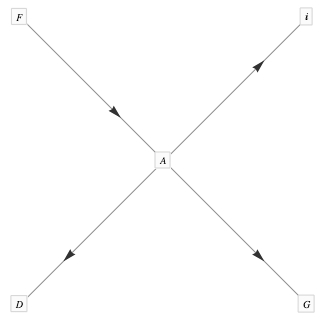}
    b)\includegraphics[width=0.5cm]{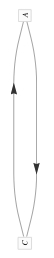}
    c)\includegraphics[width=5cm]{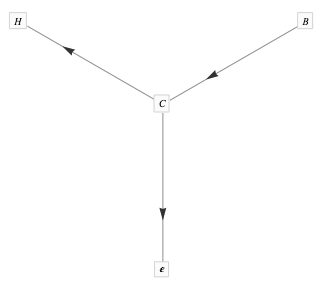}
   \caption{Decomposition of Figure \ref{fig:ximg-2}c) to molecular stars and a molecular bridge.}
   \label{fig:ximg-9}
\end{figure}

\begin{figure}[htbp]
   \centering
    \includegraphics[width=12cm]{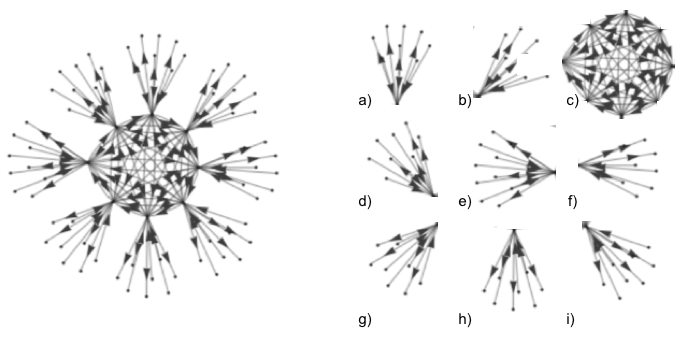}
    \caption{Decomposition to eight molecular stars with triggers and reactors and $k$ molecular diamonds.}
   \label{fig:ximg-22}
\end{figure}

\begin{figure}[htbp]
   \centering
    a) \includegraphics[width=6cm]{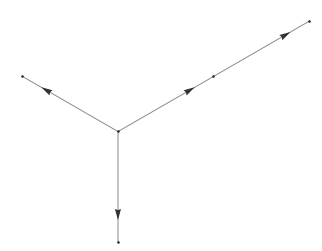}
    b) \includegraphics[width=6cm]{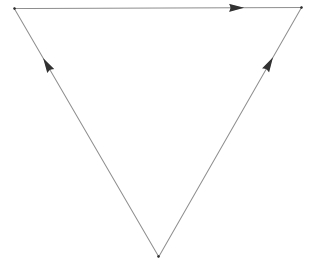}
    \caption{Examples for 3-ary molecules.}
   \label{fig:ximg-37}
\end{figure}

In contract to this, a collection of associations like $A\Rightarrow B$ and $B\Rightarrow C$ yields on a stringent sequence from nucleus A to nucleus C where the A can not be reached by B and C (as an example, see Figure \ref{fig:ximg-37} with a \textit{molecular arrow} and a \textit{molecular triangle}). In the following, we will concern only with 2-ary \textit{molecular structures}.

\section{Algorithmic Decomposition}
A decomposition is then as follows: let $a_i, a_j,$ and $a_k$ disjunctive \textit{author nuclei} with $1\leq i\neq j\neq k\leq s$ natural numbers, and \textit{STAR}, \textit{BRIDGE}, and \textit{DIAMOND} some data structures to manage identified \textit{molecule structures}. A \textit{reactor} is defined as to be true (= 1) for a nucleus when an association to another nucleus exists, otherwise false (= 0):

\begin{center}
   \textbf{reactorp($a_i$)} = $\left\{
                          \begin{array}{ll}
                          1 : & \exists a_j$ with $a_j \rightarrow a_i \\
                          0 : & else\\
                         \end{array}
                         \right.$
\end{center}

A \textit{trigger} is defined as to be true for a nucleus when an association to another nucleus exists, otherwise false:

\begin{center}
   \textbf{triggerp($a_i$)} = $\left\{
                          \begin{array}{ll}
                          1 : & \exists a_j$ with $a_i \rightarrow a_j \\
                          0 : & else\\
                         \end{array}
                         \right.$
\end{center}

Furthermore, the \textit{single atomic bonds} is a predicate that evaluates to true (= 1), when exactly one directed connection exists between two \textit{author nuclei} (i.e., the exclusive OR):

\begin{center}
   \textbf{sbond($a_i, a_j$)} = $\left\{
                          \begin{array}{ll}
                          1 : & a_i \rightarrow a_j \vee a_j \rightarrow a_i \\
                          0 : & else\\
                         \end{array}
                         \right.$
\end{center}

A \textit{double atomic bond} is a predicate evaluates to true (= 1), when \textit{single atomic bonds} exist between both $a_i, a_j$ and $a_j, a_i$, respectively:

\begin{center}
   \textbf{dbond($a_i, a_j$)} = $\left\{
                          \begin{array}{ll}
                          1 : & a_i \rightarrow a_j \wedge a_j \rightarrow a_i \\
                          0 : & else\\
                         \end{array}
                         \right.$
\end{center}

The predicate emptyp$(a_i)$ evaluates to true when $a_i$ shares no bonds (neither single nor double):

\begin{center}
   \textbf{emptyp($a_i$)} = $\left\{
                          \begin{array}{ll}
                          1 : & \not\exists a_j: a_i \rightarrow a_j \vee a_j \rightarrow a_i\\
                          0 : & else\\
                         \end{array}
                         \right.$
\end{center}

Moreover, the decomposition to either a \textit{molecular star}, \textit{molecular bridge}, or \textit{molecular diamond} can be discovered by simple functions that base on the previously defined predicates. In this respect, a \textit{molecular star} exists for a number of \textit{author nuclei} $a_1,\dots,a_s$ in case that the function STAR$(a_i)$ leads to a positive result.

\begin{verbatim}
computeSTAR ( a_i )
for all 1 <= j <= s do:
      if ( a_j exists with sbondp(a_i, a_j) )
         then add a_j to STAR (a_i)
      od;
 exit with STAR (a_i)
\end{verbatim}

STAR($a_i$) reads in an arbitrary \textit{author nucleus} and exits with a complete list of single bonds associated with $a_i$. The average computation time is polynomial. A \textit{molecular bridge} exists, when the function BRIDGE($a_i)$ is non-empty. 

\begin{verbatim}
computeBRIDGE (a_i)
for all 1 <= j <= s do:
      if ( a_j exists with dbondp(a_j, a_i) )
         then add a_j to BRIDGE (a_i)
      od;
exit with BRIDGE(a_i)
\end{verbatim}

BRIDGE($a_i$) reads in an arbitrary \textit{author nucleus} and exits with a list of bridges for each of them. As for STAR, the average computation time is polynomial. Finally, a \textit{molecular diamond} exists when the function DIAMOND($a_i$) is non-empty, returning a list of \textit{author nuclei} associating $a_i$. 

\begin{verbatim}
computeDIAMOND (a_i)
for all 1 <= j, k <= s do:
     if ( BRIDGE (a_i, a_j), BRIDGE (a_j, a_k), BRIDGE (a_k, a_i) )
         then add (a_j, a_k) to DIAMOND( a_i )
    od;
exit with DIAMOND (a_i)
\end{verbatim}

\section{Social Networking}

With the defined predicates and functions we are then able to decompose \textit{molecular structures}. In this sense, \textit{molecular stars} can be seen communities that consist of an arbitrary number of triggers and reactors; and a \textit{molecular diamond} is nothing else than a composition of bridges. Furthermore, a decomposition of \textit{molecular structures} can then be performed quite easily, leaving to a number of descriptive attributes like shown in Table \ref{tab:attributes}.

\subsection{Social Network Clustering with/without envelopes}
\begin{figure}[htbp]
   \centering
   \includegraphics[width=12cm]{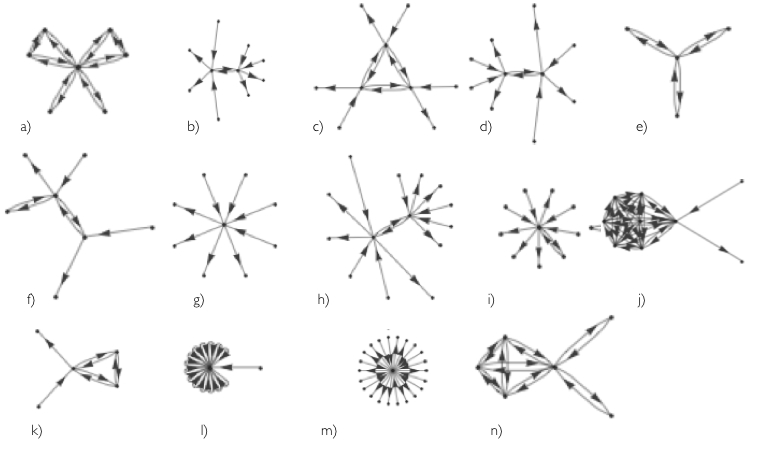} 
   \caption{Selected molecular structures that appear in visualizing associative communities.}
   \label{fig:ximg-27}
\end{figure}

Figure \ref{fig:ximg-27} briefly displays a selection of molecular structures that have appeared over the years. For structure \ref{fig:ximg-27}a), we notice a mixture of \textit{molecular diamonds} and \textit{molecular stars}, for structure \ref{fig:ximg-27}c) a mixture of \textit{molecular bridges} and \textit{atomic bonds}, and for structure \ref{fig:ximg-27}d) three bridges only, concentrated on one node. All of the other structures are similar and decomposable. The different roles to be observed fully correspond to the ones given above: triggers (outgoing link) and reactors (incoming link).

Using such a data table for clustering, we may then get groups of social sub-networks being similar. This is a simplification of existing molecular communities. For example, while taking the raw attributes data SB (number of single bonds), BR (number of bridges), DI (number of diamonds), NU (number of nuclei), RE (number of reactor nodes), and TR (number of triggers nodes) (\ref{tab:attributes}) for a hierarchical clustering, compute the distances between these attributes individually, we may group the examples following their similarities and therefore perform a clustering. 

\begin{table}[htdp]
\begin{center}
\begin{tabular}{||c|c|c|c|c|c|c||}
\hline
Mol & SB & BR & DI & NU & RE & TR\\
\hline
\ref{fig:ximg-4}a) & 7 & 0 & 0 & 8 & 1 & 7\\
\hline
\ref{fig:ximg-4}b) & 7 & 0 & 0 & 8 & 7 & 1\\
\hline
\ref{fig:ximg-4}c) & 7 & 0 & 0 & 8 & 4 & 4\\
\hline
\ref{fig:ximg-2}a) & 0 & 3 & 1 & 3 & 3 & 3\\
\hline
\ref{fig:ximg-2}b) & 5 & 3 & 1 & 8 & 7 & 4\\
\hline
\ref{fig:ximg-2}c) & 7 & 1 & 0 & 8 & 4 & 7\\
\hline
\ref{fig:ximg-2}d) & 0 & 6 & 4 & 4 & 4 & 4\\
\hline
\end{tabular}
\end{center}
\caption{Description of Molecular Structures from \ref{fig:ximg-4}a) to \ref{fig:ximg-4}c) and \ref{fig:ximg-2}a) to \ref{fig:ximg-2}d).}
\label{tab:attributes}
\end{table}

With this decomposition to \textit{n-ary molecules}, we demand on decomposing each publishing community and to describe a \textit{publishing community} by the molecular attributes. Applying such a data table containing a description for molecular structures with clustering, we may then get groups of molecular structures being similar. The advantage of such an analytical performance is a simplification of existing molecular communities in respect to their structure.

\subsection{Social Role Discovery}
\begin{figure}[htbp]
   \centering
   \includegraphics[width=12cm]{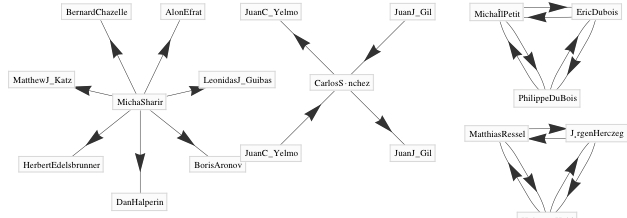} 
   \caption{Selected Social Communities in 1994.}
   \label{fig:ximg-38}
\end{figure}

The immediate identification of roles in social communities is shown in Figure \ref{fig:ximg-38}: here, we may observe \textit{molecular diamonds} and  \textit{molecular stars}, having \textit{Micha Sharir} as \textit{molecular trigger} for seven other authors. Furthermore, \textit{Carlos Sanchez} is both a \textit{molecular trigger} and a \textit{molecular reactor}, whereas \textit{Eric Dubois}, \textit{Phillipe Dubois}, and  \textit{Michael Petit} form a  \textit{molecular diamond}.

\begin{figure}[htbp]
   \centering
   \includegraphics[width=12cm]{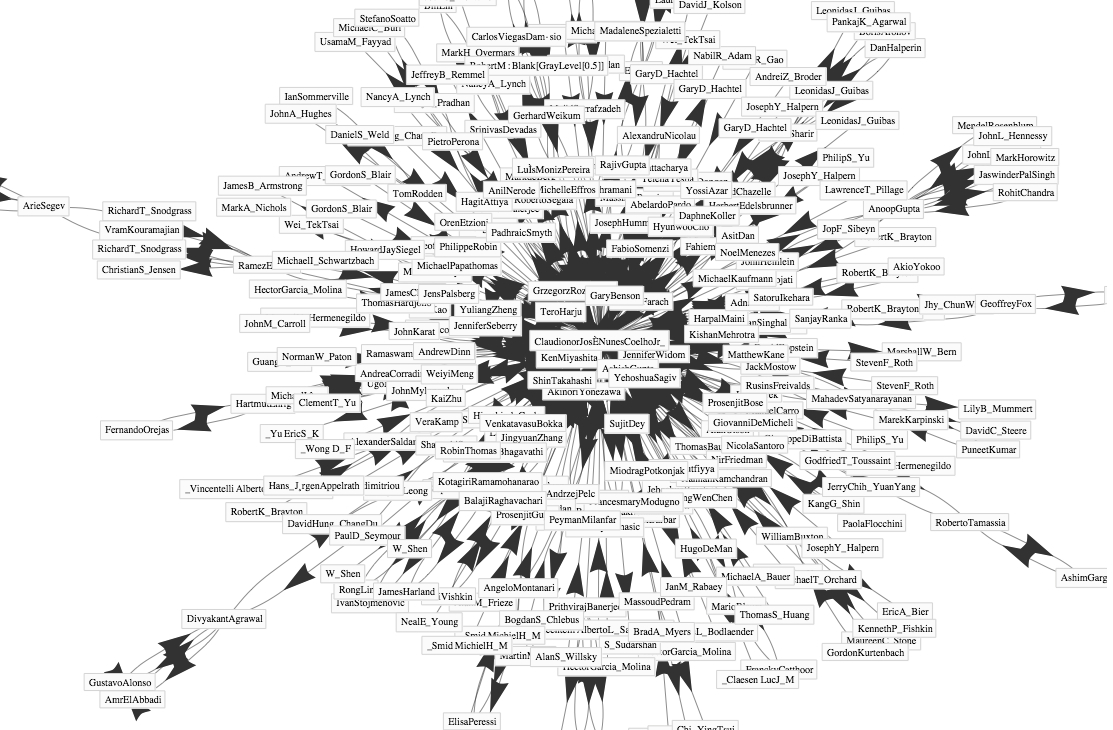} 
   \caption{Social Community exampling the big star.}
   \label{fig:ximg-42}
\end{figure}

A more complex scenario can be observed in \ref{fig:ximg-42}, which references to the \textit{big star}. This \textit{big star} is often present over many years, referencing a huge community sharing publications. For this, this structure symbolizes fruitful years with strong interrelations and cooperations.

We may observe that the \textit{big star} is nothing else than a composition of many \textit{author nuclei} in its inner centre, being connected to external \textit{nuclei}. Moreover, we may observe these \textit{nuclei}, being either a \textit{reactor}, a \textit{trigger} or both. Overall, the molecule is composed of different levels, representing an increasing importance, the more concentrated a nucleus is positioned. One nucleus in the core has much more a trigger function than a node outside, and the influence concerning a directing of research might be stronger, if we admit that this can be influenced by publications.

\subsection{Temporal Comparison of Social Networks}
Initially, we observe very simple \textit{molecules} in the years before 1950, because less publications have been made. The first \textit{molecular bridge} can be observed in 1953, the first more complex structure in 1954. The evolvement remembers to cell division operations of natural processes, leading to a first \textit{big star} in 1960. Interestingly, the \textit{molecular noise} (pairwise, but disjunctive publication, not sharing publications with others) is present the whole time, continuously staying on a similar percentage. Furthermore, the years of e.g. 1961, 1975, 1978, 1991, and 1993 are of specific notice, as they do not hold a \textit{big star}, but signifying even a evolutionary step from one research topic to another. Moreover, the latest years of 1995 to 2005 appear continuously without a dominating molecules, which might be the result of an evolving internet with lots of potential in many areas. And especially this vivid and colorful landscape enforces the scientific community to a multitude of research activities, finally shown in publications to different areas. On the other side, the amount of human researchers might be grown up, the possibilities of electronically publish a work has been tremendously increased. 

\begin{figure}[htbp]
   \centering
    \includegraphics[width=12cm]{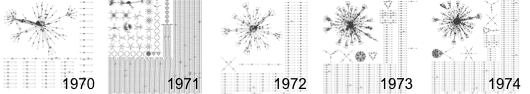}
    \includegraphics[width=12cm]{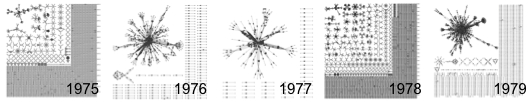}
    \includegraphics[width=12cm]{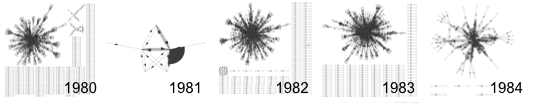}
    \includegraphics[width=12cm]{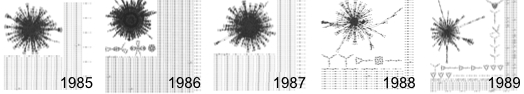}
    \includegraphics[width=12cm]{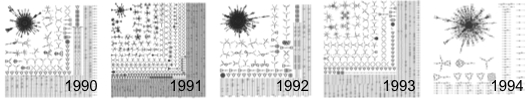}
    \includegraphics[width=12cm]{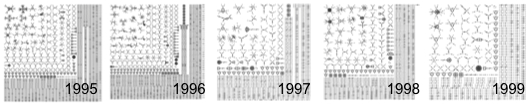}
    \caption{Selected molecular communities, showing the years from 1970 to 1999.}
   \label{fig:ximg-40}
\end{figure}

A yearly flight over the association landscape between 1970 and 1999 yields on a results as presented in Figure \ref{fig:ximg-40}. The first years are characterized by an alternating appearance of the \textit{big star} (two consecutive years) and one year of restructuring. This is, for example, in 1975, 1978, and 1981. Interestingly, the research years where \textit{Artificial Intelligence} had become significantly could be characterized by the social communities between 1982 and 1990, dominating the publication landscape with less space for other social communities. In contrast to this, the social communities in the 1990's not generally concern with one social domain but stay manifold and distributed, sharing more simple \textit{molecular structures} than in the years before.

\section{Conclusions}

\begin{figure}[htbp]
   \centering
    \includegraphics[width=12cm]{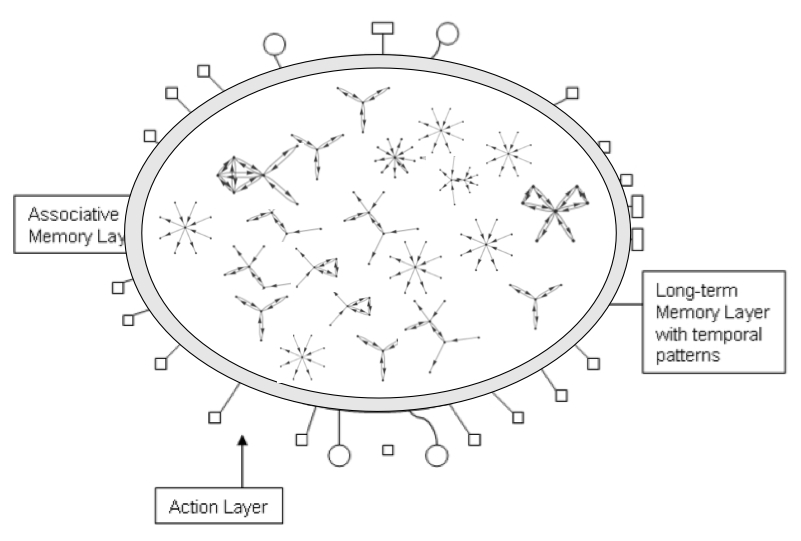}
    \caption{Social Communities in the associative memory layer: of an incremental-adaptive Mind-map.}
   \label{fig:ximg-44}
\end{figure}

We have focused on entries of the bibliographic communities DBLP and characterized communities through a simple typifying description model. We have set a publication as a transaction between its associated authors, the general idea is to concern with directed associative relationships amongst them, to decompose each pattern to the fundamental \textit{molecular components}, and to describe these communities by such \textit{atomic} and \textit{molecular} attributes. The decomposition supports the management of discovered structures towards the use of adaptive-incremental mind-maps (Figure \ref{fig:ximg-44}), being discovered \textit{molecular} structures at the \textit{associative memory layer} and firstly managed in the \textit{short-term} memory.
Understanding bibliographic entries as data stream input, this is an important step towards the interpretation of (temporal) \textit{social communities} as informational and intermediate results. 

\section*{Acknowledgement}
This work has been done in the scope of the research project ICC, which is currently performed at the MINE research group. MINE is member of the ILIAS Computer Science Laboratory of the Department of Computer Science at the University of Luxembourg.

\end{document}